\documentclass[prl,superscriptaddress,twocolumn,showpacs]{revtex4}
\usepackage{graphicx}
\usepackage{times}
\usepackage{amsmath,amssymb}
\newcommand{\be}{\begin{equation}}
\newcommand{\ee}{\end{equation}}
\newcommand{\bea}{\begin{eqnarray}}
\newcommand{\eea}{\end{eqnarray}}

\newcommand{\zl}{|0\rangle_l}
\newcommand{\onel}{|1\rangle_l}
\newcommand{\zlh}{|\widehat{0}\rangle_l}
\newcommand{\onelh}{|\widehat{1}\rangle_l}
\newcommand{\Phat}{\widehat{P}}

\begin{document}

\title{Fibonacci topological order from quantum nets}

\author{Paul Fendley}
\affiliation{Department of Physics, University of Virginia, Charlottesville, VA 22904-4714 USA}

\author{Sergei V. Isakov}
\affiliation{Theoretische Physik, ETH Zurich, 8093 Zurich, Switzerland}

\author{Matthias Troyer}
\affiliation{Theoretische Physik, ETH Zurich, 8093 Zurich, Switzerland}

\date{\today} 

\begin{abstract} 

We analyze a model of quantum nets and show it has non-abelian topological order of doubled Fibonacci type. The ground state has the same topological behavior as that of the corresponding string-net model, but our Hamiltonian can be defined on any lattice, has less complicated interactions, and its excitations are dynamical, not fixed. This Hamiltonian includes terms acting on the spins around a face, around a vertex, and special ``Jones-Wenzl'' terms that serve to couple long loops together. We provide strong evidence for a gap by exact diagonalization, completing the list of ingredients necessary for topological order.
\end{abstract} 
\maketitle
\paragraph{Introduction--}

The theoretical and experimental search for condensed-matter systems with fractionalized excitations 
has attracted considerable attention. One reason is the promise of protection against errors in a topological quantum computer \cite{Kitaev97,Nayakreview}. Another is more fundamental: fractionalization provides a dramatic example of how rich the emergent long-distance behavior of a many-body system can be.  

Finding magnetic systems with fractionalized excitations, in particular those with the non-abelian statistics necessary for topological quantum computation, is a difficult but important problem.  Major progress resulted from the introduction of the ``toric code'' or ``quantum-double'' models \cite{Kitaev97}, and the ``string-net'' models \cite{LevinWen}.  These Hamiltonians are sums of commuting projectors, each of which annihilates the ground state. Since all can be diagonalized, the full gapped spectrum of the Hamiltonian is easily computed. The excitations are effectively immobile defects and so are non-dynamical. Even though these solvable models are quite special, the presence of the gap means they exemplify a particular phase; the topological order necessary for fractionalized excitations persists around the exactly solvable points  \cite{Klich,Bravyi}.

In this paper we focus on a particularly important phase, that with ``Fibonacci'' topological order. This is the simplest universal topological order: any unitary operator can be approximated to arbitrary precision by quasiparticle braiding. We find a Hamiltonian with doubled Fibonacci order with several desirable characteristics not present in the corresponding string-net model \cite{LevinWen}. While the ground state is essentially the same, the excitations here are dynamical. Moreover, the interactions between the spins are simpler to describe, making the connection of the terms in the Hamiltonian with the ground-state wave function more explicit. This allows our model to be defined on any lattice, and  further illuminates the topological order. To construct it, we expand upon Ref.\ \onlinecite{Fendley08} by including terms in the Hamiltonian with the essential property of coupling loops in different winding sectors. By extensive numerical exact diagonalization, we then provide strong evidence that the resulting Hamiltonian is indeed gapped.  

\paragraph{Quantum loops and nets--}
 {Quantum loop models} provide an elegant picture of the way fractionalized excitations arise in a magnetic model \cite{Freedman01}. Fractionalized excitations in a magnetic model are typically attached by a defect line or ``string''. The string itself costs no energy, because away from the quasiparticles, it is locally indistinguishable from the ground state itself. The existence of the string can however result in anyonic statistics,  because rotating one quasiparticle around another causes the strings to interact, even when the quasiparticles are far apart.  The Hilbert space is spanned by non-intersecting string configurations, and the Hamiltonian includes an energy penalty for each ``string end''. Thus the ground state can include only states containing closed loops. Any state with a string end then corresponds to a quasiparticle excitation, and the defect lines are the strings attaching the quasiparticles. 
More picturesquely, ``cutting'' a loop results in an anyon pair. 
There are thus for example three distinct configurations with four quasiparticles, corresponding to how the strings attach them. Since away from a quasiparticle a string with ends is indeed indistinguishable from a closed loop, these three configurations should have the same energy. Braiding can cause a transition between the different configurations, and possibly result in non-abelian statistics.

Unfortunately, non-abelian statistics does not occur in quantum loop models in their simplest formulation  \cite{FNS,Trebst06,Troyer08}. A way of salvaging the idea is to modify the inner product to have more desirable topological properties \cite{Fendley08}. This seemingly innocuous change has dramatic consequences:  in the new orthonormal basis, the excitations are naturally described in terms of {\em quantum nets}, configurations with branching. This naturally leads both to an elegant quantum self-duality and to the Hilbert space of the string-net models!

\paragraph{The vertex terms in the Hamiltonian--}
Our Hamiltonian is of ``Rokhsar-Kivelson'' \cite{RK} or ``frustration-free'' type. It is the sum over (not-commuting) projectors, such that there exists at least one state annihilated by each of the projectors. Such a state is necessarily a ground state, and all ground states must be annihilated by all projectors. Quasiparticle excitations then necessarily involve configurations not annihilated by at least one of these projectors; the lowest-energy excitations are those dominated by configurations where only a few of the projectors acting on it do not annihilate it. 

We study models with a two-state orthonormal quantum system $\zl$, $\onel$ at each {\em link} $l$ of some two-dimensional lattice. The basis elements of the Hilbert space are then the tensor product of the states for each link.
A convenient way of picturing each basis state is to draw a line on each link $l$ when the state $|1\rangle_l$ is present, while leaving the link empty for $|0\rangle_l$. This way, each basis state is pictured by some geometric object; the strings are formed from links with $|1\rangle_l$.

A quantum net Hamiltonian includes an energy penalty for ``string ends'', to ensure none appear in the ground state \cite{Kitaev97,Freedman01,LevinWen}. A string end at  {vertex} $v$ is a configuration with one link $l$ touching $v$ having the state $\onel$ and the others with $|0\rangle$. Let the projector $P_l$ act as
$P_l  (\cdots \otimes \zl\otimes\cdots) =  (\cdots \otimes \zl\otimes\cdots) $ and $
P_l  (\cdots \otimes \onel\otimes\cdots) = 0\ .
$
Thus on the square lattice, with $v_1,v_2,v_3,v_4$ the links touching $v$, the operator
\bea
\nonumber
H_v &=& (1-P_{v_1})P_{v_2}P_{v_3}P_{v_4}+(1-P_{v_2})P_{v_1}P_{v_3}P_{v_4}\\
&&+\  (1-P_{v_3})P_{v_1}P_{v_2}P_{v_4} +(1-P_{v_4})P_{v_1}P_{v_2}P_{v_3}
\eea
gives $1$ and $0$ respectively on any basis state with and without a string end at $v$. Adding $\sum_v H_v$ to a  Hamiltonian comprising sums of projectors therefore forbids string ends in a zero-energy ground state. 
The zero-energy state
$|\Psi\rangle$ annihilated by  $\sum_v H_v$ can thus be written as a sum over ``nets'' $N$,
geometric objects with no ends:
\begin{equation}
|\Psi\rangle = \sum_N w(N) |N\rangle\ ,
\label{PsiN}
\end{equation}
Other terms in the Hamiltonian then will determine the values of the weights $w(N)=\langle N|\Psi\rangle $ in the ground state.

\paragraph{A ground state with non-abelian topological order--}

The weights $w(N)$ of the configurations in the ground state (\ref{PsiN}) must satisfy certain properties to ensure topological order. For example, the number of ground states depends on the genus of the surface on which the model is defined; for doubled Fibonacci topological order
there is only one ground state when space is topologically a sphere, but four
for a torus \cite{Fidkowski}.  Ground states of the form (\ref{PsiN}) naturally exhibit this behavior, because 
geometric objects like loops and nets have winding numbers around cycles of a surface with non-trivial topology. Multiple ground states occur when the local terms in the Hamiltonian do not change these winding numbers.

A gap in the spectrum is also desirable. While a gap is a property of the Hamiltonian, the ground state of any gapped local Hamiltonian in two dimensions satisfies an important constraint: all expectation values of {\em local} operators decay exponentially as the operators are moved apart \cite{Hastings}. On the flip side, however, to have deconfined anyons, the expectation value of some {\em non-local} operators must decay {algebraically}; geometric objects like loops or nets should have long-range correlations. Exponential decay here implies a vanishing probability for two would-be fractionalized excitations attached by a string to be far apart, so they are confined.

To have abelian topological order, a ground state satisfying these properties is typically sufficient. To have fractionalized excitations with non-abelian statistics, however, much more structure need be present. Multi-quasiparticle states must be degenerate so that braiding can cause a transition to another state. However, this is not enough \cite{FNS,Trebst06,Troyer08}; the inner product must also have the appropriate topological properties \cite{Fendley08}. 

Non-abelian topological order can occur when the topological part of ground-state weight $w(N)$  is given by a {\em chromatic polynomial} \cite{FF}. The chromatic polynomial $\chi_{\widehat{N}}(Q)$ is easiest to understand by treating the strings in the net $N$ as borders separating countries. The dual graph $\widehat{N}$ corresponds to a vertex for each country and an edge connecting each pair of countries sharing a border. For $Q$ integer, $\chi_{\widehat{N}}(Q)$ is the number of ways of coloring each country with $Q$ colors such that neighboring countries (i.e.\ two adjacent vertices on $\widehat{N}$)  are colored differently. It is a polynomial in $Q$ that can be evaluated for $Q$ non-integer as well, and by definition is a topological invariant. Any string end results in $\chi_{\widehat N} (Q)\,=\,0$ because then ${\widehat N}$ has a vertex attached to itself.

We focus on the simplest example of universal non-Abelian topological order, the ``Fibonacci'' case $Q=\phi^2=\phi+1$, where $\phi = (1+\sqrt{5})/2$, the golden ratio. The ground-state weight $w(N)$ of any net model with (doubled) Fibonacci order, including the string-net model \cite{LevinWen}, must necessarily involve the chromatic polynomial \cite{Fidkowski}. This topological order and the resulting excitations have been discussed in depth in \cite{Fidkowski,Burnell}.  The degrees of freedom in the Fibonacci string-net model \cite{LevinWen} are a two-state system $\zl$, $\onel$ on the links of the honeycomb lattice, and the unnormalized ground state $|\Psi\rangle$ in (\ref{PsiN})  is summed over all net configurations with \cite{Fidkowski}
\begin{equation}
w(N) =w_s(N)\equiv \phi^{3t_N/4}
\chi_{\widehat{N}}(\phi^2)\ ,
\label{wstringnet}
\end{equation}
where $t_N$ is the number of trivalent vertices in the net $N$ (i.e.\ those with all three neighboring links in the state $|1\rangle$).  To find our Hamiltonian, we study a ground state slightly different than (\ref{wstringnet}). Following \cite{Fendley08}, we take
\be
w(N) = \phi^{-L_N/2} \chi_{\widehat{N}}(\phi^2)\ ,
\label{wus}
\ee
where the ``length'' $L_N$ of the net $N$ is the number of links it covers, i.e.\ the number of states $\onel$. Since the weights (\ref{wstringnet},\ref{wus}) are identical in their topological properties, it is natural to expect that the latter results in the same doubled-Fibonacci topological order as the string-net ground state (\ref{wstringnet}). Numerical and analytical arguments indicate that indeed such a ground state has all the desired properties described above \cite{FJ,Fendley08}.

\paragraph{The face terms in the Hamiltonian--}

The remainder of this paper is devoted to finding a gapped Hamiltonian annihilating the ground state $|\Psi\rangle$ in (\ref{PsiN}) with weights (\ref{wus}).  An advantage of using the weights (\ref{wus}) is that they are ``quantum self-dual'': they take the same form when rewritten on the dual lattice \cite{Fendley08}. This results in projectors annihilating this ground state involving only links around a face of the lattice, as opposed to the 12-spin interaction in the string-net model \cite{LevinWen}. Moreover, these projectors are easily defined on any lattice.

Seeing this requires making a change of basis from $\zl,\onel$ on each link  to another basis $\zlh,\onelh$ via the unitary matrix $F=F^{-1}$ defined by
\be
\begin{pmatrix}
\zlh\\
\onelh
\end{pmatrix}
=
\frac{1}{\phi}
\begin{pmatrix}
1 &\sqrt{\phi}\\
\sqrt{\phi} & -1 
\end{pmatrix}
\begin{pmatrix}
\zl\\
\onel
\end{pmatrix}\ .
\ee
Not coincidentally, this matrix is also the fusion matrix for four Fibonacci anyons. The quantum self-duality stems from interpreting this new basis as describing nets on the {\em dual} lattice. The links of the dual lattice are in one-to-one correspondence with those of the original lattice, so we are free to define such a net $D$ by drawing a string on the {\em dual} link when the state $\onelh$ is present, and leaving the dual link empty when $\zlh$ is present.
The remarkable fact is that when the ground state $|\Psi\rangle$ is rewritten in the new basis, one obtains \cite{FK,Fendley08}
\bea 
\label{PsiD}
|\Psi\rangle &=& \sum_D \widehat{w}(D) |D\rangle\\
\label{wD}
\widehat{w}(D)&=& \langle D |\Psi\rangle = \alpha\,
\phi^{-L_D/2} \chi_{\widehat{D}}(\phi^2) \ ,
\eea
where $L_D$ is the length of the net $D$, i.e.\ the number of dual links with $\onelh$, while $\alpha$ is an unimportant constant. Comparing (\ref{PsiD},\ref{wD}) with (\ref{PsiN},\ref{wus}) shows the weighting of the nets of the dual lattice is completely equivalent to the weighting on the original lattice, hence the quantum self-duality.

In particular, (\ref{wD}) means that any configuration $D$ with string ends has $\widehat{w}(D)=0$ and does not appear in the ground state. Thus a projector $H_f$ for vertices on the dual lattice, analogous to $H_v$ on the original lattice, will also annihilate this ground state. The links $f_1$, $f_2,\dots $ touching a vertex on the dual lattice correspond to links around a {\em face} on the original lattice. Thus defining the projector $\Phat_l = F P_l F$ onto $\zlh$ means that for each original face on the square lattice, the projector
\bea
H_f &=& (1-\Phat_{f_1})\Phat_{f_2}\Phat_{f_3}\Phat_{f_4}+ (1-\Phat_{f_2})\Phat_{f_1}\Phat_{f_3}\Phat_{f_4}\\
&&+\  (1-\Phat_{f_3})\Phat_{f_1}\Phat_{f_2}\Phat_{f_4}+(1-\Phat_{f_4})\Phat_{f_1}\Phat_{f_2}\Phat_{f_3}
\nonumber
\eea
annihilates $|\Psi\rangle$. This operator is non-diagonal in the original net basis and so couples different net configurations. 

\begin{figure}[t]
\centerline{
\includegraphics[width=\columnwidth]{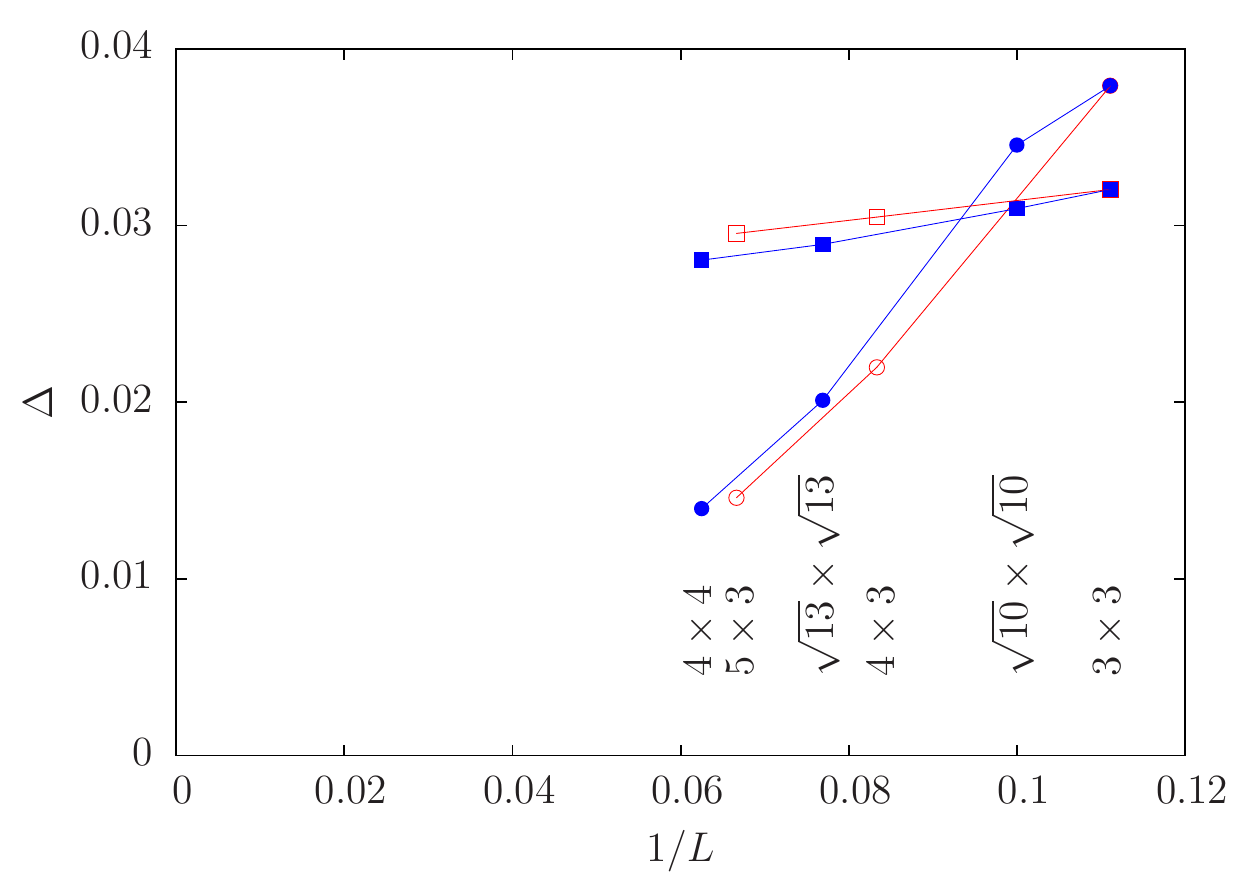}
}
\caption{The gap to the first excited state as a function of the inverse system size at $\varepsilon=0$ (circles, no Jones-Wenzl terms) and at $\varepsilon=0.8$ (squares). Lines guide the eye. Extrapolation to $L\to\infty$ indicates the system is gapless at $\varepsilon=0$ and gapped at $\varepsilon=0.8$.}
\label{fig:gap2}
\end{figure}

The Hamiltonian $H_{\rm vf}=\sum_v H_v + \sum_f H_f$ thus annihilates the desired ground state $|\Psi\rangle$, and only has interactions around each vertex and face, so e.g.\ only has four-spin interactions for the square lattice. The last remaining thing to check is if it has a gap. Our numerical results unfortunately indicate that there is {\em not} a gap for this Hamiltonian, see fig.~\ref{fig:gap2}.
Another  (probably related) problem is that $H_{\rm vf}$ does not have the right number (four) of ground states
for doubled Fibonacci topological order on the torus. The number grows with the size of the torus, which follows by rewriting $|\Psi\rangle$ in a (non-orthonormal) loop basis; the unwanted ground states correspond to loops wrapping around the cycles of the torus \cite{Fendley08}. 

\paragraph{Coupling wrapping loops via Jones-Wenzl terms---}

A virtue of writing the ground-state weights in terms of topological quantities like the chromatic polynomial is that one can find many local projectors annihilating it \cite{FF}. An important such projector is the ``Jones-Wenzl'' projector \cite{JW}. When added to the Hamiltonian, it not only couples the long loops and so removes the unwanted ground states, but can give a gap \cite{Freedman01}. For example, the toric code Hamiltonian can be rewritten in the form $H_{\rm vf}$ plus Jones-Wenzl type terms appropriate for $Q=2$ \cite{Fendley08}. We here derive these terms for the Fibonacci ground state (\ref{wus}), and provide strong evidence using numerical exact diagonalization that the spectrum now includes a gap.

\begin{figure}[h] 
\begin{center} 
\includegraphics[width= .35\textwidth]{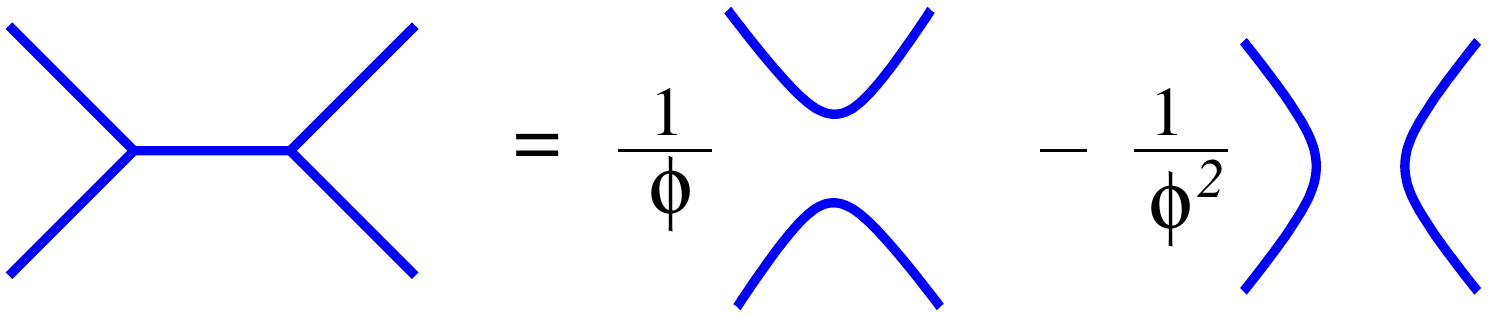} 
\caption{The Jones-Wenzl identity for $Q=\phi^2$}
\label{fig:jw} 
\end{center}
\end{figure}
\begin{figure}[h] 
\begin{center} 
\includegraphics[width= .4\textwidth]{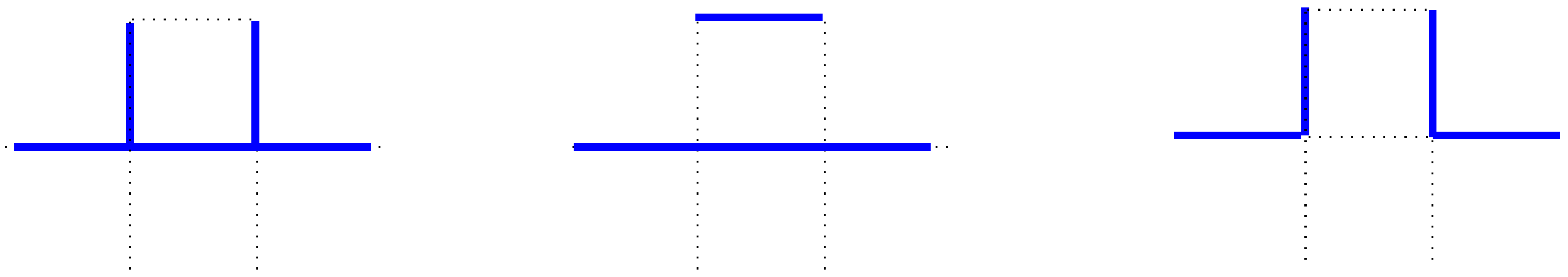} 
\caption{Three configurations mixed by the Jones-Wenzl projector}
\label{fig:3configs} 
\end{center}
\end{figure}

 
The Jones-Wenzl projectors for knot polynomials result in an identity for each chromatic polynomial at $\cos^{-1}(\sqrt{Q/2})/\pi$ rational \cite{FK}, which for
$Q=\phi^2$ is \cite{Tutte}
\begin{equation}
\chi_{\widehat{t}} (\phi^2)= 
\frac{1}{\phi} \chi_{\widehat{E}}(\phi^2) 
- \frac{1}{\phi^2} \chi_{\widehat{I}}(\phi^2)\ 
\label{jw}
\end{equation}
with the three nets $t,E,I$ involved illustrated in fig.\ \ref{fig:jw}. This identity is true {\em locally}, meaning it holds for any portion of a net. 
Consider the three nets $|t\rangle$, $|E\rangle$, and $|I\rangle$ 
on the square lattice displayed in fig.\ \ref{fig:3configs}, identical
everywhere except on one face.  Because the ground state (\ref{wus}) is written in terms of chromatic polynomials,
the identity (\ref{jw}) results in a local relation between amplitudes in
the ground-state wavefunction:
\begin{equation}
\langle t|\Psi\rangle =
{\phi^{-3/2}}\, \langle E|\Psi\rangle -
{\phi^{-5/2}}\, \langle I|\Psi\rangle\ .
\label{jw2}
\end{equation}
The extra factor of $\sqrt{\phi}$ in (\ref{jw2})
relative to (\ref{jw}) results from the weight per unit length in
(\ref{wus}). 
\vskip-.17in

\begin{figure}[h] 
\begin{center} 
\includegraphics[width= .49\textwidth]{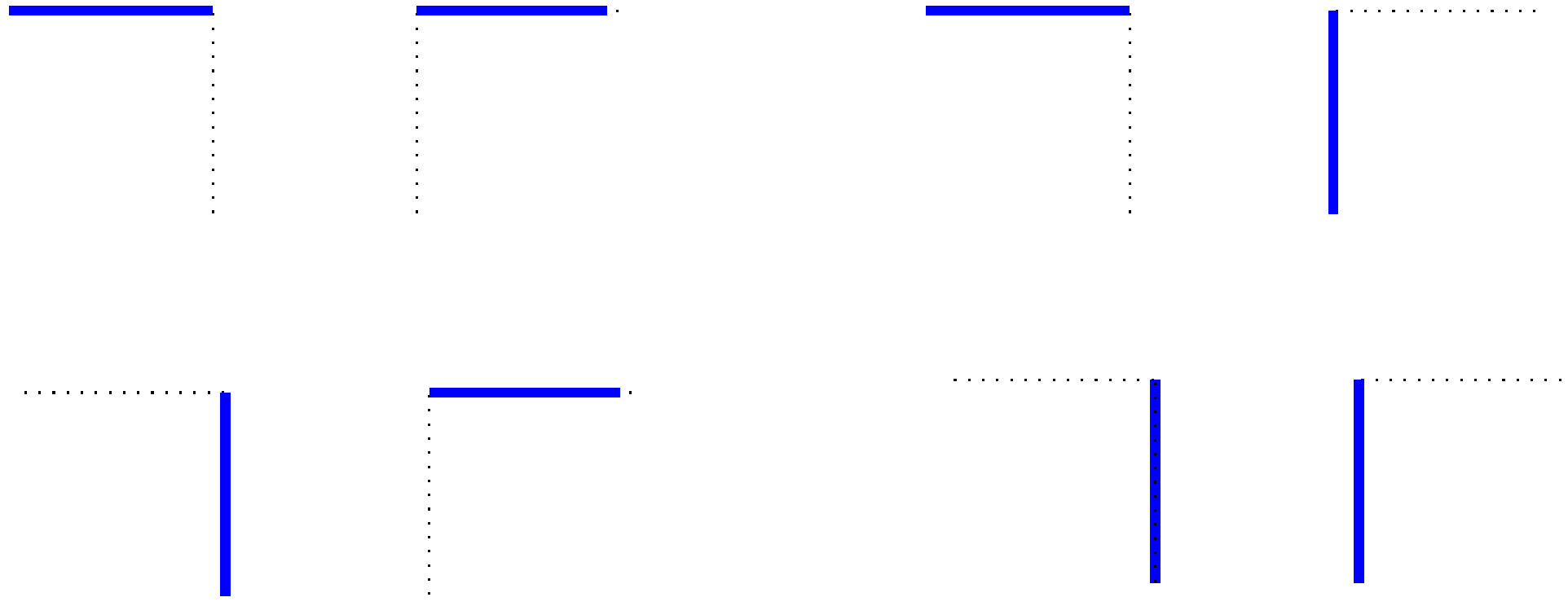} 
\caption{Configurations on the outside links not annihilated
  by $J_{\rm out}$}
\label{fig:outsidelinks} 
\end{center}
\end{figure} 

We exploit the relation (\ref{jw2}) to find a  projector $H_j$ that annihilates the
ground state $|\Psi\rangle$ and couples the winding loops. 
It is of the form $H_j = J_{\rm in} J_{\rm out}$, and acts non-diagonally on the four ``inside'' links around the square in fig.\ \ref{fig:3configs}, and diagonally on the remaining four
``outside'' links. The non-diagonal $J_{in}$ projects onto  the linear combination $\phi^{5/2} |t\rangle  - \phi |E\rangle + | I\rangle $. A Hermitian operator doing this is
\begin{equation}
{J}_{\rm in}=
\begin{pmatrix}
\phi^{5/2}& -\phi &1\\
-\phi& \phi^{-1/2}& -\phi^{-3/2}\\
1&-\phi^{-3/2}&\phi^{-5/2}
\end{pmatrix}
\end{equation}
acting on $|t\rangle$, $|E\rangle$ and $|I\rangle$ respectively; it annihilates all other configurations on the inner links. The diagonal part $J_{\rm out}$  annihilates any configuration on the outer links other than those illustrated in fig.\ 
\ref{fig:outsidelinks}, i.e.\ labeling the outside links $j_1$, $j_2$, $j_3$ and $j_4$ from left to right and letting $Q_j = (1-P_j)$ gives 
\bea
\nonumber
J_{\rm out} &=& Q_{j_1}P_{j_2} Q_{j_3} P_{j_4}\ +\  Q_{j_2}P_{j_1} Q_{j_3} P_{j_4}\\
 &&
\ +\ Q_{j_1}P_{j_2} Q_{j_4} P_{j_3} + Q_{j_2}P_{j_1} Q_{j_4} P_{j_3}\ .
\eea
Each 
$H_j$ therefore mixes each of four sets of three configurations among themselves; one such set is illustrated in fig.\ 
\ref{fig:3configs}, while the other three are given by changing the
configurations on the outer links to one of the others in
fig.\ \ref{fig:outsidelinks}. Because of $J_{\rm out}$,  $H_j$ does not mix net configurations
with non-net configurations. Using (\ref{jw2}) then ensures that $H_j|\Psi\rangle =0$ for $j$ any set of eight links of the form illustrated in fig.\ \ref{fig:3configs}, i.e.\ the four spins on a face and the four touching two adjacent vertices. There are thus four such terms for each face on the lattice.  Since $|\Psi\rangle$ can be rewritten in terms of nets on the dual lattice as well, the analogous projectors $H_{\widehat j}$ also annihilating $|\Psi\rangle$ can be defined by repeating the above arguments acting on the $\zlh,\onelh$ basis on the dual lattice. The full Hamiltonian is then
\be
H = \sum_v H_v + \sum_f H_f + \varepsilon \left( \sum_j H_j + \sum_{\widehat j} H_{\widehat j} \right),
\label{eq:hamiltonian}
\ee
where $\varepsilon$ is a coupling strength included for convenience.

\begin{figure}[t]
\centerline{
\includegraphics[width=\columnwidth]{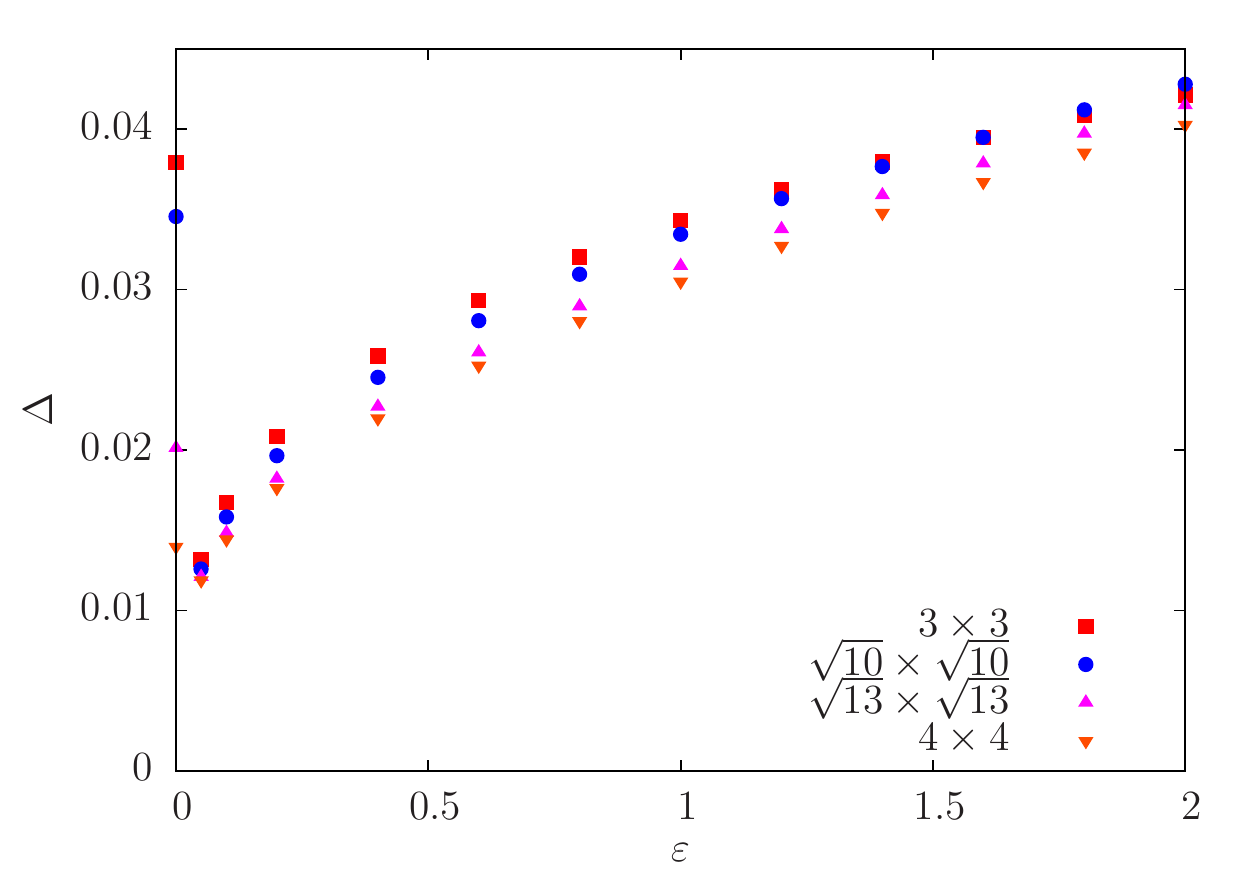}
}
\caption{The gap to the first excited state as a function $\varepsilon$ for different system sizes.}
\label{fig:gap1}
\end{figure}

\paragraph{Numerical results---} We have confirmed by exact diagonalization that the Hamiltonian (\ref{eq:hamiltonian}) has exactly four ground states when $\epsilon>0$, as required for doubled Fibonacci topological order. More importantly, we have also checked that $H$ is gapped when $\epsilon>0$. 
We diagonalized $H$ by the Lanczos method, using a parallel and highly scalable exact diagonalization code \cite{ed}. 
The $2L_1L_2$ spins live on the links of a $L_1\times L_2$ square lattice with periodic boundary conditions. Clusters of size $3\times3$, $\sqrt{10}\times\sqrt{10}$, $4\times3$, $\sqrt{13}\times\sqrt{13}$, $5\times3$, and $4\times4$ were diagonalized, all  but the $4\times3$ and $5\times 3$ having the symmetry of the infinite lattice. We exploit translation symmetry to reduce the Hilbert-space size, so the largest Hilbert space size is about $2.7\times10^8$ for the $4\times4$ lattice. It is not very large, but the complicated multi-spin interactions result in many matrix elements per row; e.g.\ roughly 16000 for the $4\times4$ lattice.  
The lowest eigenvalues for $\varepsilon > 0$ are in the zero-momentum sector.
In fig.~\ref{fig:gap1}, we show the gap $\Delta$ to the first excited state as a function of $\varepsilon$ for different system sizes. To illustrate how the gap survives in the thermodynamic limit, we show in fig.~\ref{fig:gap2} its finite-size scaling. This suggests the gap is indeed non-zero, with an estimate of $\Delta\approx 0.025$ in the thermodynamic limit for $\varepsilon=0.8$. We checked as well the gap including the $H_j$ but omitting the dual projectors $H_{\widehat j}$, and obtained about half this value.

\paragraph{Conclusion--}
We have found a gapped magnetic model with doubled Fibonacci topological order which can be constructed on any two-dimensional lattice. We believe this is the ``minimal'' gapped Hamiltonian with an exact ground state in this phase acting on a two-state ``spin'' Hilbert space. Moreover, the excitations are dynamical, and not fixed defects.  Since the connection of the terms in the Hamiltonian with the ground state is apparent in our construction, this provides useful insight into finding even-more-physical Hamiltonians with the same topological order. To this end, it would be interesting to find an analogous two-body Hamiltonian acting on a Hilbert space with more states per site \cite{OT}.
It would also be illuminating to apply a similar analysis to other models with non-abelian topological order. A good starting point may be the models of e.g.\ Refs.\ \onlinecite{Chamon,Paredes}, related to the one considered here. Another interesting topic for further study is the presumed quantum critical point at $\epsilon=0$ describing a transition out of the topological phase; an analogous point also occurs in the correspondingly deformed toric code.

\paragraph{Acknowledgments--}
The work of P.F.\ is supported by the US National Science Foundation under the grant DMR/MPS1006549, while  the work of S.I.\ and M.T.\ is supported by  the Swiss HP2C initiative.
M.T.\ acknowledges hospitality of the Aspen Center for Physics, supported by 
NSF grant \#1066293.
The calculations were performed with the MAQUIS-ED code, developed with
support of the HP2C, on the Brutus cluster at ETH Zurich and
the Cray XE6 ``Monte Rosa'' at CSCS.

\def\cmp#1#2#3{Comm.\ Math.\ Phys.\ {\bf #1}, #2 (#3)}

\end{document}